\documentclass[12pt]{article}

\usepackage{a4}

\begin{document}
\title{Dirac-Yang monopoles and their regular counterparts}
\author{{\large Tigran Tchrakian}$^{\dagger \star}$ \\ \\ $^{\dagger}${\small Department of Mathematical Physics, National University of Ireland Maynooth,} \\ {\small Maynooth, Ireland} \\ \\ $^{\star}${\small Theory Division, Yerevan Physics Institute (YerPhI), AM-375 036 Yerevan 36, Armenia}}

\date{}
\newcommand{\dd}{\mbox{d}}
\newcommand{\tr}{\mbox{tr}}
\newcommand{\la}{\lambda}
\newcommand{\ka}{\kappa}
\newcommand{\al}{\alpha}
\newcommand{\ta}{\theta}
\newcommand{\ga}{\gamma}
\newcommand{\Ga}{\Gamma}
\newcommand{\de}{\delta}
\newcommand{\si}{\sigma}
\newcommand{\Si}{\Sigma}
\newcommand{\f}{\phi}
\newcommand{\vf}{\varphi}
\newcommand{\F}{\Phi}
\newcommand{\bomega}{\mbox{\boldmath $\omega$}}
\newcommand{\bsi}{\mbox{\boldmath $\sigma$}}
\newcommand{\bchi}{\mbox{\boldmath $\chi$}}
\newcommand{\bal}{\mbox{\boldmath $\alpha$}}
\newcommand{\bpsi}{\mbox{\boldmath $\psi$}}
\newcommand{\brho}{\mbox{\boldmath $\varrho$}}
\newcommand{\beps}{\mbox{\boldmath $\varepsilon$}}
\newcommand{\bxi}{\mbox{\boldmath $\xi$}}
\newcommand{\bbeta}{\mbox{\boldmath $\beta$}}
\newcommand{\ee}{\end{equation}}
\newcommand{\eea}{\end{eqnarray}}
\newcommand{\be}{\begin{equation}}
\newcommand{\bea}{\begin{eqnarray}}
\newcommand{\ii}{\mbox{i}}
\newcommand{\e}{\mbox{e}}
\newcommand{\pa}{\partial}
\newcommand{\Om}{\Omega}
\newcommand{\vep}{\varepsilon}
\newcommand{\bfph}{{\bf \phi}}
\newcommand{\lm}{\lambda}
\def\theequation{\arabic{equation}}
\renewcommand{\thefootnote}{\fnsymbol{footnote}}
\newcommand{\re}[1]{(\ref{#1})}
\newcommand{\R}{{\rm I \hspace{-0.52ex} R}}
\newcommand{\N}{{\sf N\hspace*{-1.0ex}\rule{0.15ex}%
{1.3ex}\hspace*{1.0ex}}}
\newcommand{\Q}{{\sf Q\hspace*{-1.1ex}\rule{0.15ex}%
{1.5ex}\hspace*{1.1ex}}}
\newcommand{\C}{{\sf C\hspace*{-0.9ex}\rule{0.15ex}%
{1.3ex}\hspace*{0.9ex}}}
\newcommand{\eins}{1\hspace{-0.56ex}{\rm I}}
\renewcommand{\thefootnote}{\arabic{footnote}}

\maketitle
\begin{abstract}
The Dirac-Yang monopoles are singular Yang--Mills field configurations
in all Euclidean dimensions. The regular counterpart of the Dirac
monopole in $D=3$ is the t~Hooft-Polyakov monopole, the former being
simply a gauge transform of the asymptotic fields of the latter. Here, regular
counterparts of Dirac-Yang monopoles in all dimensions, are described.
In the first part of this talk the hierarchy of Dirac--Yang (DY) monopoles
will be defined, in the second part the motivation to study these in
a topoical context will be briefly presented, and in the last part,
two classes of regular counterparts to the DY hierarchy will be presented.
\end{abstract}
\medskip
\medskip
%\newpage

\section{The Dirac--Yang hierarchy in $D\ge 3$}
The Dirac~\cite{D} monopole can be constructed by gauge transforming the asymptotic
't~Hooft-Polyakov monopole~\cite{T-P} in $D=3$, which can be taken to be spherically
symmetric~\footnote{It is not infact necessary to restrict to spherically symmetric
fields only. By choosing to start with the asymptotic axially symmetric fields
characterised with vorticity $n$, the gauge tranformed connection is just $n$
times the usual Dirac monopole field.}, such that the $SO(3)$ isovector
Higgs field is gauged to a (trivial) constant, and the $SU(2)\sim SO(3)$
gauge group of the Yang-Mills (YM) connection breaks down to
$U(1)\sim SO(2)$, the resulting Abelian connection developing a line
singularity on the positive or negative $(x_3=)z$-axis.

In exactly the same way, the Yang~\cite{Y} monopole can be constructed
by gauge transforming the asymptotic $D=5$ dimensional 'monopole'~\cite{mono} such that
the such that the $SO(5)$ isovector Higgs field is gauged to a (trivial) constant, and
the $SO(5)$ gauge group of the YM connection breaks down to
$SO(4)$, the resulting non Abelian connection developing a line
singularity on the positive or negative $x_5$-axis. In fact, the residual non
Abelian connection can take its values in one or other chiral representations of
$SU(2)$, as formulated by Yang~\cite{Y}, but this is a low dimensional accident which
does not apply to the higher diemnsional analogues to be defined below, all of which
are $SO(D-1)$ connctions. Just like the
't~Hooft-Polyakov monopole monopole is the regular counterpart of the Dirac monopole,
so is the $D=5$ dimensional 'monopole'~\cite{mono} the regular counterpart of the
Yang monopole.

The above two definitions of the Dirac and of the Yang monopoles will
be the template for our definition of what we will refer to as the
hierarchy of Dirac--Yang (DY) monopoles in all dimensions. The two
examples just given are both in odd ($D=3$ and $D=5$) dimensions, but
the DY hierarchy is in fact defined in all, including even, dimensions.

Just as the Dirac monopole can be defined as a gauge transform of the
asymptotic spherically symmetric 't~Hooft-Polyakov monopole,
our definition for the DY fields in arbitrary $D$ dimensions starts from the
(non Abelian) $SO(D)$ YM field $A_i$ and the $D$-tuplet Higgs field $\F$
\bea
A_i^{(\pm)}=\frac{1}{r}\,\Si_{ij}^{(\pm)}\,\hat x_j\quad&,&\quad
\F=\hat x_i\,\Si_{i,D+1}^{(\pm)}\qquad,\qquad{\rm for\ odd}\ \ D\label{YMHodd}\\
A_i^{(\pm)}=\frac{1}{r}\,\Ga_{ij}\,\hat x_j\quad&,&\quad
\F=\hat x_i\,\Ga_{i,D+1}\qquad,\qquad{\rm for\ even}\ \ D\,.\label{YMHeven}
\eea
$\hat x_i=\frac{x_i}{r}$, $i=1,2,..,D$, is the unit radius vector. $\Ga_i$ are the
Dirac gamma matrices in $D$ dimensions with the chiral matrix $\Ga_{D+1}$ for even
$D$, so that
\[
\Ga_{ij}=-\frac{1}{4}\left[\Ga_i,\Ga_j\right]
\]
are the Dirac representations of $SO(D)$. The matrices $\Si_{ij}$, employed only in
the odd $D$ case, are
\[
\Si_{ij}^{(\pm)}=-\frac{1}{4}\left(\frac{\eins\pm\Ga_{D+2}}{2}\right)
\left[\Ga_i,\Ga_j\right]\,,
\]
$\Ga_{D+1}$ being the chiral matrix in $D+1$ dimensions, and $\Si_{ij}^{(\pm)}$
being one or other of the two possible chiral representations of the $SO(D)$
subgroup of $SO(D+1)$. 

That \re{YMHodd}-\re{YMHeven} are the asymptotic fields of regular monopoles in
$D$ dimensions is the subject of the last part of this talk, while in the next
part we will argue why such regular finite energy monopoles are relevant. Here,
we define the DY field configurations as gauge transforms of
\re{YMHodd}-\re{YMHeven}.

The DY monopoles result from the action of the following $SO(D)$ gauge group
element
\be
\label{g}
g_{\pm}=\frac{(1\pm\cos\ta_1)\eins\pm\Ga_D\Ga_{\al}\hat x_{\al}\sin\ta_1}
{\sqrt{2(1\pm\cos\ta_1)}}\,,
\ee
having parametrised the $\R^D$ coordinate $x_i=(x_{\al},x_D)$ in terms of the radial
variable $r$ and the polar angles
\be
\label{polar}
(\ta_1,\ta_2,..,\ta_{D-2},\vf)
\ee
with the index alpha running over
$\al=1,2,..,D-1$.  The meaning of the $\pm$ sign in \re{g} is as follows: Choosing
these signs the Dirac line sinularity will be along the negative or positive
$x_D$--axis, respectively. (In the case of odd $D$ if we chose the opposite sign
on $\Si$ in \re{YMHodd} the situation will be reversed.) In other words the DY
field will be the $SO(D-1)$ connection on the upper or lower half $D-1$
sphere, $S^{D-1}$, respectively, the transition gauge transformation being given
by $g_+g_-^{-1}$. Notice that the dimensionality of the matrices $g$, \re{g}, and
those of both \re{YMHodd} and \re{YMHeven}, match in each case.

In $D>3$ dimensions, the gauge group element \re{g} was first employed in \cite{OT}
and \cite{AOT} in $D=4$, and was subsequestly extended to all $D$ in \cite{MT} and
\cite{TZ}.

The result of the action of \re{g} on \re{YMHodd} or \re{YMHeven},
\bea
A_i\rightarrow &&gA_ig^{-1}+g\pa_ig^{-1}\nonumber\\
\F\rightarrow &&g \F g^{-1}\nonumber
\eea
yields the required DY fields
$\hat A_i^{(\pm)}=(\hat A_{\al}^{(\pm)},\hat A_D^{(\pm)})$
\bea
\hat A_{\al}^{(\pm)}&=&\frac{1}{r(1\pm\cos\ta_1)}\,\Si_{\al\beta}
\hat x_{\beta}\quad,\quad\hat A_{D}^{(\pm)}=0\quad,\quad{\rm for\ odd}\ D
\label{DYodd}\\
\hat A_{\al}^{(\pm)}&=&\frac{1}{r(1\pm\cos\ta_1)}\,\Ga_{\al\beta}
\hat x_{\beta}\quad,\quad\hat A_{D}^{(\pm)}=0\quad,\quad{\rm for\ even}\ D\,,
\label{DYeven}
\eea
and the Higgs field is gauged to a constant, i.e. it is trivialised.

The components of the DY curvature
$\hat F_{ij}^{(\pm)}=(\hat F_{\al\beta}^{(\pm)},\hat F_{\al D}^{(\pm)})$
follow from \re{DYodd}-\re{DYeven} straightforwardly. To save space we give only the
curvature corresponding to \re{DYodd}
\bea
\hat F_{\al\beta}^{(\pm)}&=&-\frac{1}{r^2}\left[\Ga_{\al\beta}+
\frac{1}{(1\pm\cos\ta_1)}\,\hat x_{[\al}\,\Ga_{\beta]\ga}\hat x_{[\ga}\right]
\label{fab}\\
\hat F_{\al D}^{(\pm)}&=&\pm\frac{1}{r^2}\,\Ga_{\al\ga}\hat x_{\ga}\,,\label{fad}
\eea
where the notation $[\al\beta]$ implies the antisymmetrisation of the indices, and
the components of the curvature for even $D$ corresponding to \re{DYeven} follows
by replacing $\Ga$ in \re{fab}-\re{fad} with $\Si^{(\pm)}$.
The parametrisation \re{DYodd}-\re{DYeven} and \re{fab}-\re{fad} for the DY field
appeared in \cite{MT} and \cite{TZ}.

That the DY field \re{DYodd}-\re{DYeven} in $D$ dimensions, constructed by
gauge transforming the asymptotic fields \re{YMHodd}-\re{YMHeven} of a $SO(D)$ EYM
system, is a $SO(D-1)$ YM field is obvious. For $D=3$ and $D=5$, these are the
Dirac~\cite{D} and Yang~\cite{Y} monopoles, respectively.

In retrospect, we point out that to construct DY monopoles it is not even necessary
to start from a YMH system, but ignoring the Higgs field and simply applying the
gauge transformation \re{g} to the YM members of \re{YMHodd}-\re{YMHeven} results
in the DY monopoles \re{DYodd}-\re{DYeven}. In other words the only function of the
Higgs fields in \re{YMHodd}-\re{YMHeven} is the definition of the gauge group
element \re{g} designed to trivialise it.

We will henceforth restrict our detailed considerations concerning the regular
counterparts of the DY monopoles, to the first two lowest dimensions, namely $D=3$
and $D=4$. This excludes even the Yang monopole itself, but it is more instructive
since we then deal both with an odd and and even $D$. Before that however, we will
motivate briefly the role of the regular monopoles in the next part of this talk.

\section{Motivation}
Field theory solitons in higher dimensions find application~\cite{Tong} as the
$D$-branes of string theory, and also, for open heterotic
strings~\cite{Pol} in the absence of gravity. As solitons of string theory, $D$-branes
must be finite energy/mass solutions of the appropriate gravitating field theories.

When non Abelian matter gravitates, there occur both regular and black hole
solutions with finite mass/energy, in contrast with Abelian matter where only
black hole solutions exist.
In $3+1$ spacetime dimensions the gravitating YM field,
both in the absence~\cite{BK} and in the presence~\cite{W,BFM} of the isovector
Higgs field has been intensively studied. The Dirac monopole field features in
these solutions as a limiting field configuration in the form of
Reissner--Nordstr{\o}m (RN) solutions of the Einstein--Maxwell system.

In $D+1$ spacetime dimensions, with $D\ge 4$, the gravitating YM field again has
both regular~\cite{BCT} and black hole~\cite{BCHT} solutions with finite mass/energy.
The situation is the same also in the presence of a negative~\cite{RT} and a
positive~\cite{BRT} cosmological constant. Again, higher dimensional 
Reissner--Nordstr{\o}m (RN) solutions appear as limiting solutions~\cite{BMT}, but
the latter feature non Abelian gauge fields now, unlike in the $D=3$ case where
the gauge sector of the RN field is the usual, Abelian, Maxwell field.
These are the DY monopoles introduced above.

The fields DY \re{DYodd}-\re{DYeven} and \re{fab}-\re{fad} appeared recently in \cite{GT},
where it was shown that they satisfy the gravitating YM equation
(for the usual $p=1$ YM system) is satisfied by them in all dimensions $D$,
with or without cosmological constant. This
is not surprising, since {\it in the presence of gravity} the second order field
equations to YM systems consisting of the superposition of all possible members of
$p$-hierarchy (defined below by \re{YMhier}) are satisfied by DY fields.

In \cite{BCT,BCHT} we have
constructed finite mass solutions to the $(p=1)+(p=2)$ YM model in $D=4,5,6,7$, or
spacetimes $d=5,6,7,8$ for the spherically symmetric $SO(D)$ YM connection
\be
\label{sph}
A_i=\frac{1-w(r)}{r}\,\Si_{ij}\,\hat x_j\,.
\ee
Setting the function $w(r)=0$ by hand reduces \re{sph} to the singular Wu-Yang (WY)
part of the field \re{YMHodd}-\re{YMHeven}, which we know are gauge equivalent to the
DY fields \re{DYodd}-\re{DYeven} and hence equations satified by the WY fields
are also satisfied by the DY fields. This result carries through to the full
superposed YM hierarchy in any given dimension, subject to satisfying finite energy
scaling requirements. Of course when gravitating YM solutions are constructed, $w(r)=0$
is not set by hand. These are the DY fields which arise as the RN
configurations for as limiting solutions~\cite{BMT}.

There remains to see what the interesting properties of the gravitating WY
(with $w=0$ in \re{sph}) fields, are. Clearly, these have to be black hole
solutions since the WY fields are singular at the origin. In \cite{RT} we have
given the mass function $m(r)$ (first member of Eqn (24) therein) for the field \re{sph}
in arbitrary dimensions, for the gravitating YM system consisting of the full
superposition of $p$-YM terms. In the WY limit, i.e. with $w=0$, this is
\be
\label{m'}
m'=\sum_{p=1}^{P}\frac{\tau_p}{2 (2p-1)!} \frac{(d-3)!}{(d-2[p+1])!}
\ r^{-(4p-d+2)}\,,
\ee
where $d=D+1$ is the dimension of the spacetime.
Obviously the mass, namely the integral of \re{m'}, will diverge for certain
combinations of $p$ and $d$. Most importantly, for $d\ge 5$ (i.e. for ``higher
dimensions'') the usual $p=1$ YM term will result in infinite mass, and for the mass
to be finite the least nonlinear YM term must be the $p=2$ one. Thus, restricting to
the usual YM term as in \cite{GT} leads to infinite mass!

In \cite{GT} it is commented that the advantage of employing singular DY
(or alternatively WY as seen above) solutions is, that they are evaluated in closed
form, unlike the regular gravitating matter solutions as e.g. \cite{BK} in $D=3$
and \cite{BCT} in $D\ge 4$. To retain this feature -- of closed form black hole
gravitating non Abelian matter solutions -- and to have finite mass, the
appropriate $p$-YM rather than (usual) $p=1$-YM terms must be employed.

Strictly speaking, for the purposes of picking out the correct $p$-YM terms in
\re{m'}, there is no need to start from the full theory that supports
regular finite energy topologically stable counterparts of the DY monopoles. One
could simply consider the (singular) black hole solution featuring the DY fields
\re{DYodd}-\re{DYeven}, or even more directly the corresponding Wu-Yang fields
\cite{MT,TZ,WY}.

\section{The regular counterparts of DY fields}
Regular solutions to gravitating non Abelian (YM) matter fall into two main classes.
The first of these is simply the solutions to the models described by the Lagrangians
consisting of the superposition of
(possibly) all members of the YM hierarchy~\footnote{The YM hierarchy of $SO(4p)$
gauge fields in the chiral
(Dirac matrix) representations consisting only of the $p$-YM term in \re{YMhier} was
introduced in \cite{hier} to construct selfdual instantons
in $4p$ dimensions. (The selfduality equation for the $p=2$ case was solved
indepenently in \cite{GKS}, whose authors subsequently stated in their {\it Erratum},
that this solution was the instanton of the
$p=2$ member of the hierarchy introduced earlier in~\cite{hier}.) The instantons of the
generic system \re{YMhier}, while stable, are not selfdual and cannot be evaluated in closed
form and are constructed numerically~\cite{BT}. Restricting ourselves here to
finite action (instanton) solutions only, it is worth mentioning an altenative hierarchy which
supports selfdual instantons in $4p+2$ dimensions~\cite{S,F}. While it is straightforward to construct spherically symmetric
solutions with gauge group $SO(4p+2)$ in the chiral Dirac representations, these selfduality equations are even more overdetermined than those of
the $4p$ dimensional hierarchy. The action densities of
these systems are not positive definite so that, while the selfduality equations do solve
the second order field equations, they do not saturate a Bogomol'nyi bound and hence are
not necessarily stable. }
\cite{hier}
\begin{equation}
\label{YMhier}
{\cal L}_P=\sqrt{-\mbox{det}g}\sum_{p=1}^{P}\frac{\tau_p}{2(2p)!}\ \mbox{Tr}\,F(2p)^2\,,
\end{equation}
$F(2p)$ denoting the $p$ fold totally antisymmetrised product
\[
F(2p)\equiv F_{\mu_1\mu_2...\mu_{2p}}=F\wedge F\wedge...\wedge F\ ,\qquad
p\ \ {\rm times}\ ,
\] of the YM
curvature, $F(2)=F_{\mu\nu}$, in this notation. Clearly, the highest value $P$
of $p$ in \re{YMhier} is finite and depends on the dimensionality $d=D+1$ of
the spacetime. To complete the definition of the models \re{YMhier}, the
gauge group $G$ must be specified. With our aim in the present paper, of constructing
static spherically symmetric solutions in $d=D+1$ spacetime dimensions, the
smallest such gauge group is $G=SO(d-1)=SO(D)$.

To \re{YMhier} is added some
gravitational Lagrangian, e.g. Einstein--Hilbert or Gauss--Bonnet, or a
superposition of these, or possibly
even a dilaton term.
Many such studies \cite{BCT,BCHT,RT,BMT} were carried out and the regular solutions
were constructed. In \cite{BMT} in particular it was pointed out that the $SO(2)$
Reissner--Nordstr{\o}m fixed point occuring in $d=3+1$ has its $SO(D-1)$
analogues for all $D$. These are indeed the DY monopole fields discussed in part
{\bf 1} above, although in \cite{BMT} we did not use that nomenclature, referring
to these simply as RN fields. Before proceeding to the second class of models, we end
our discussion of the present class by pointing out that the finite mess/energy
solutions they support do not {\it always} survive the decoupling of gravity, e.g.
in the $d=4$ case~\cite{BK}.

The second class of models consists of YM fields, {\it viz.} \re{YMhier}, interacting
with scalar matter. By far the most prominent of these are the
gauged Higgs (YMH) models~\footnote{There are other gravitating YM--scalar
matter models, e.g. the gauged Grassmannian model in $d=5$~\cite{BRT2}.} whose solitons
are stabilised by monopole charges. In $D=3$ these are the celebrated
't~Hooft--Polyakov monopoles and in $D$ dimensions those defined in Refs~\cite{mono},
which will be illustrated below, in the non gravitating case. All these models
feature an $D$--component isovector Higgs field which is instrumental (but not
essential) in our definition of DY fields in part {\bf 1}. The main difference of the
solutions of gauged Higgs systems from those of \re{YMhier} without Higgs fields is, that
they {\it always} survive the decoupling of gravity. 

While the Dirac monopole~\cite{D} and the Yang monopole~\cite{Y} are defined in
$D=3$ and $D=5$, here we will choose the dimensions $D=3$ and $D=4$ for our
illustrations, with the purpose of displaying both an odd $D$ and an even $D$
example. Even in these restictive catchment, there are two ways of constructing
YMH models. The first of these is via the dimensional reduction of $p$-YM systems
on a product space $\R^{D}\times S^{4p-D}$, while the second one is more {\it ad hoc}
and it relies on the fact that the topology of a YMH system is encoded in the Higgs
field exclusively~\cite{AFG,TZ}. The relation between these two procedures was
explored in some detail in \cite{Matin} so we give just a summary here. In both procedures,
the all important quantities are the {\it topological charges}, for whose definitions we
refer to \cite{Matin}, which enable the statment of
Bogomol'nyi inequalities leading to the $D$ dimensional models. In the first case these
are the magnetic monopole charges descending from the $2p$-th Chern--Pontryagin (CP)
charge defined on $\R^{D}\times S^{4p-D}$, while in the second case the topological charges
are the winding numbers of the Higgs field, suitably re-expressed so that the winding
numbers are the integrals of {\it gauge invariant} densities.

We will first consider the descended CP topological charge case, and then the covariantised
winding number case, for $D=3$ and $D=4$. In each cae we will define the charge density,
followed by the resulting models whose solutions support regular monopoles.

In any given dimensions $D$ the descended CP density can be constructed from any $p$-YM
system on any $\R^{D}\times S^{4p-D}$. Naturally, the examples we give here are the simplest
possibilities, pertaining to smallest possible choice for this $p$.
Descending from the $2$-nd CP density on $\R^{D}\times S^{4-D}$ and the
$4$-th CP density on $\R^{D}\times S^{8-D}$, for $D=3$ and $D=4$ respectively, the two
reduced CP (or magnetic charge) densities~\cite{OT,AOT} are
\bea
\varrho^{(3)}_{\rm CP}&=&\frac{1}{16\pi}\,\vep_{ijk}\,\mbox{Tr}\,F_{ij}\,D_k\F\quad,\quad
i=1,2,3\label{CP3}\\
&=&\frac{1}{16\pi}\,\vep_{ijk}\,\pa_k\,\mbox{Tr}\,F_{ij}\,\F\label{td3}\\
\varrho^{(4)}_{\rm CP}&=&\frac{1}{64\pi^2}\ \vep_{ijkl}\,\mbox{Tr}\,\Ga_5\bigg[
S^2F_{ijkl}+4\{S,D_i\F\}\{F_{[jk},D_{l]}\F\}\nonumber\\
&+&3\left(\{S,F_{ij}\}+\left[D_{i}\F,D_{j}\F\right]\right)
\left(\{S,F_{kl}\}+\left[D_{k}\F,D_{l}\F\right]\right)
\bigg]\quad,\quad
i=1,2,3,4\label{CP4}\\
&=&\frac{1}{64\pi^2}\ \vep_{ijkl}\,\pa_i\mbox{Tr}\,\Ga_5\bigg[\eta^4A_j\left(F_{kl}-
\frac23A_kA_l\right)
+\frac16\eta^2\F\{F_{[jk},D_{l]}\F\}\nonumber\\
&&\qquad\qquad\qquad\qquad
+\frac16\F\,\left(\{S,F_{kl}\}+\left[D_{k}\F,D_{l}\F\right]\right)D_j\F\bigg]\label{td4}
\eea
where $F_{ijkl}=\{F_{i[j},F_{kl]}\}$ is the curvature $4$-form, and we have used the notation
$S=\left(\eta^2-\F^2\right)$.

In passing, \re{td3} and \re{td4} demonstrate the fact that the topological current of a
reduced CP charge density is {\it gauge invariant} for odd $D$, and is {\it gauge variant}
for even $D$.

In this case the resulting action/energy density that supports regular finite action/energy
topologically stable solutions follows {\it uniquely} from the same dimensional descent that
yielded the charge densities \re{CP3}-\re{CP4}, now applied to the action density of the
$p$-YM system on $\R^{D}\times S^{2p-D}$ (with $p=1$ for $D=3$ and $p=2$ for $D=4$). The
descended Bogomol'nyi inequalities can be saturated only in the $p=1$ case, so that the solutions
in question are only those to the second order field equations for $p\ge 2$.

The energy/action densities bounded from below by \re{CP3}-\re{CP4}, with this bound actually
saturated in the $D=3$ case, are
\bea
{\cal S}^{(3)}&=&\frac14\,\mbox{Tr}\left(F_{ij}^2+2\,D_i\F^2\right)\label{s3}\\
{\cal S}^{(3)}&=&\frac{1}{48}\mbox{Tr}\left(F_{ijkl}^2+4\{F_{[jk},D_{l]}\F\}^2+18
\left(\{S,F_{ij}\}+\left[D_{i}\F,D_{j}\F\right]\right)^2+5\{S,F_{kl}\}^2+54\,S^4
\right).\label{s4}
\eea
The DY gauge here has a particularly enlightening application. In this gauge, all Higgs dependent
terms in \re{s3}-\re{s4} vanish and all we are left with are the $2$-form and $4$-form YM terms.
What is more is that this shows that the asymptotic behaviour of any of these monopoles is
such that the curvatrure $2$-form decays as $r^{-2}$, unlike instantons.

It may be interesting here to remark that in $D=4$, where we performed the
descent over $\R^{4}\times S^{4}$ yielding \re{CP4} and \re{s4}, we could
have opted instead to descend over the six dimensional
space $\R^{4}\times S^{2}$. In that case the appropriate six dimensional YM system~\footnote{Departing from our brief for
a moment and considering a monopole in $D=5$ on the other hand, it is indeed possible to
descend from a purely $p=2$ YM term on $\R^{5}\times S^{1}$, so residual system in this case
would feature only a $F_{ijkl}^2$ term with a valid topological lower bound~\cite{mono}.}
would have been
\[
\mbox{Tr}\left(\frac14\,F_{\mu\nu}^2+\frac{\ka}{48}\,F_{\mu\nu\rho\si}^2\right)\,,
\]
if the residual action is to be bounded from below by a topological charge, in this case the
$3$rd CP charge. But then the residual model would have featured a $F_{ij}^2$ term whose
volume integral diverges by virtue of the asymptotics explained in the previous paragraph.

Next we give the suitably gauge covariantised~\cite{Matin} winding number densities in terms of
the usual winding number density
\be
\label{wn}
\varrho_0^{(D)}=\vep_{i_1i_2...i_D}\vep^{a_1a_2...a_D}
\pa_{i_1}\phi^{a_1}\pa_{i_2}\phi^{a_2}...\pa_{i_D}\phi^{a_D}\,,
\ee
which is not gauge invariant, and the gauge invariant density
\be
\label{cd}
\varrho_G^{(D)}=\vep_{i_1i_2...i_D}\vep^{a_1a_2...a_D}
D_{i_1}\phi^{a_1}D_{i_2}\phi^{a_2}...D_{i_D}\phi^{a_D}\,,
\ee
which is not a total divergence. For the purpose at hand it is more convenient to use a
component notation for the $SO(D)$ YM connection and the $D$-plet Higgs field
\[
A_i=-\frac12\,A_i^{aa'}\,\Si_{aa'}\quad,\quad\F=-\frac12\,\f^a\,\Si_{aD+1}
\]
for odd $D$, with $\Si$ replaced by $\Ga$ for even $D$.
These charge densities are,
\bea
\varrho_{\rm wind}^{(3)}&=&\varrho_0^{(3)}+\frac{1}{4\pi}\frac32\vep_{ijk}\vep^{baa'}
\pa_i\left(A_j^{aa'}\phi^b\pa_k|\phi^c|^2\right)\label{top3v}\\ &=&
\varrho_G^{(3)}+\frac{1}{4\pi}.\frac32\vep_{ijk}\vep^{baa'}
F_{ij}^{aa'}\phi^b\pa_k|\phi^c|^2\label{top3i}
\eea
for $D=3$, and for $D=4$,
\bea
\varrho_{\rm wind}^{(4)}&=&\varrho_0^{(4)}-
\pa_{i}\left(|\vec\phi|^2\pa_{j}\Omega_{ij}\right)-\frac38
\vep_{ijkl}\vep^{bb'cc'}\pa_{i}
\left\{(\eta^4-|\vec\phi|^4)A_{j}^{cc'}\left[\pa_{k}A_{l}^{bb'}+
\frac23\left(A_{\rho}A_{l}\right)^{bb'}\right]\right\}\label{top4v}\\
&=&\varrho_G^{(4)}+\frac32\vep_{ijkl}\vep^{bb'cc'}\left\{
\left(\pa_{i}|\vec\phi|^2\right)F_{kl}^{cc'}\phi^bD_{j}\phi^{b'}
+\frac{1}{16}\left(\eta^4-|\vec\phi|^4\right)
F_{ij}^{bb'}F_{kl}^{cc'}\right\}\label{top4i}
\eea
where $\Omega_{ij}$ denotes the {\it gauge variant} tensor quantity
\be
\label{O}
\Omega_{ij}=\frac32\vep_{ijkl}\vep^{bb'cc'}A_{l}^{cc'}
\phi^b\left(\pa_{k}\phi^{b'}+D_{k}\phi^{b'}\right)\,,
\ee
which vanishes when subjected to spherical symmetry irrespective of the
detailed asymptotic decay of the fields. The surface integrals of the total divergence term in
\re{td3} and \re{td4} vanish for suitable finite energy/action boundary conditions, so that
the topological charge here is simply the winding number. The Bogomol'nyi inequalities are
constructed from the gauge covariant charge densities \re{top3i} and \re{top4i}. This is quite a
straighforward procedure, but increasingly non unique with increasing dimension. The only caveat
is to exclude those possibilities not consistent with finite energy/action requirements for
a Higgs model. We will not list these here as they are not particularly instructive and rather
cumbersome, the $D=3$ case being given in \cite {Matin}. Perhaps the main distinctive feature
of energy/action densities bounded by \re{td3}-\re{td4} {\it versus} those bounded by
\re{CP3}-\re{CP4} instead is, that the energy/action of the models constructed via dimensional
descent always have smaller energy/action than those arrived at directly via winding
number considerations.

\begin{small}

\end{small}
\medskip
\medskip

\end{document}